\documentclass[conference]{IEEEtran}
\IEEEoverridecommandlockouts
\usepackage{cite}
\usepackage{amsmath,amssymb,amsfonts}
\usepackage{algorithmic}
\usepackage{graphicx}
\usepackage{textcomp}
\usepackage{xcolor}
\usepackage{url}                                
\usepackage{graphicx}                           
\usepackage{float}
\usepackage{subcaption}                         
\usepackage[small]{cwpuzzle}                    
\usepackage{amsmath}                            
\usepackage{tikz}                               
\usepackage{dirtytalk}                          
\usepackage{color}                              
\usepackage{blindtext}                          
\usepackage[utf8]{inputenc}
\usepackage[T1]{fontenc}

\def\BibTeX{{\rm B\kern-.05em{\sc i\kern-.025em b}\kern-.08em
    T\kern-.1667em\lower.7ex\hbox{E}\kern-.125emX}}

\begin{document}

\title{Problem Solving at the Edge of Chaos: \\ Entropy, Puzzles and the Sudoku Freezing Transition
\thanks{This work is partly sopported by the Brazilian Research Agencies CAPES, CNPq and FAPERGS.}
}

\author{\IEEEauthorblockN{Marcelo Prates}
\IEEEauthorblockA{\textit{Institute of Informatics} \\
\textit{Federal University of Rio Grande do Sul}\\
Porto Alegre, Brazil \\
morprates@inf.ufrgs.br}
\and
\IEEEauthorblockN{Luis Lamb}
\IEEEauthorblockA{\textit{Institute of Informatics} \\
\textit{Federal University of Rio Grande do Sul}\\
Porto Alegre, Brazil \\
lamb@inf.ufrgs.br}
}

\maketitle

\begin{abstract}
Sudoku is a widely popular $\mathcal{NP}$-Complete combinatorial puzzle whose prospects for studying human computation have recently received attention, but the algorithmic hardness of Sudoku solving is yet largely unexplored. In this paper, we study the statistical mechanical properties of random Sudoku grids, showing that puzzles of varying sizes attain a hardness peak associated with a critical behavior in the constrainedness of random instances. In doing so, we provide the first description of a Sudoku \emph{freezing} transition, showing that the fraction of backbone variables undergoes a phase transition as the density of pre-filled cells is calibrated. We also uncover a variety of critical phenomena in the applicability of Sudoku elimination strategies, providing explanations as to why puzzles become boring outside the typical range of clue densities adopted by Sudoku publishers. We further show that the constrainedness of Sudoku puzzles can be understood in terms of the informational (Shannon) entropy of their solutions, which only increases up to the critical point where variables become frozen. Our findings shed light on the nature of the $k$-coloring transition when the graph topology is fixed, and are an invitation to the study of phase transition phenomena in problems defined over \emph{alldifferent} constraints. They also suggest advantages to studying the statistical mechanics of popular $\mathcal{NP}$-Hard puzzles, which can both aid the design of hard instances and help understand the difficulty of human problem solving.
\end{abstract}


\section{Introduction}

Sudoku is a combinatorial puzzle whose origins trace back to Leonhard Euler's work with Latin Squares \cite{gomes2002completing}, but which has achieved unprecedented worldwide popularity in the last decade. Dubbed \say{the Rubik's cube of the $21^{th}$ century} \cite{pendlebury2005can}, the Sudoku phenomenon has sparked a number of mathematical and scientific studies, including using the structure of the puzzle to perform both traditional and quantum cryptography \cite{chithra2015enhancing,huggan2017sudoku,jones2016quantum} and devise cytometry devices \cite{yang2017sudoku}. The constraints of Sudoku puzzles have also been harnessed to successfully generate the first demonstration of non-Markovian light, by imprinting overlapping Sudoku solutions on spatial light modulators \cite{eichelkraut2015photonics}. Perhaps most interestingly, due to their popularity and $\mathcal{NP}$-Hardness \cite{yato2003complexity} Sudoku puzzles have found a niche in the interdisciplinary study of human learning and problem solving. Leu and Abbas \cite{leu2016computational} use Sudoku to study human problem-solving skills and their acquisition, representing them in a cognitively plausible manner with neural networks. The puzzle has also been used as a benchmark for collaborative problem-solving on social networks on more than one occasion \cite{farenzena2011collaboration,tabajara2013leveraging}, and for working memory in cognitive processing on at least one \cite{leu2014role}. Apart from mathematical results such as the minimum necessary number of clues to ensure solution uniqueness or the properties of the Sudoku symmetry group \cite{mcguire2012there,arnold2013nest}, the Sudoku fever has sparked developments and insightful analogies in the social \cite{giampietro2015analogy} and physical sciences \cite{williams2012paramagnetic,ercsey2012chaos}.

The growing interest by Sudoku inside and outside academia raises the need for an understanding of which factors contribute to make Sudoku puzzles challenging. If Sudoku is to be used as a benchmark for human reasoning and problem-solving, we must first be able to assess the hardness of Sudoku solving, but surprisingly this matter has been given little attention. \cite{pelanek2014difficulty} have raised this very concern, to which they propose a computational model of human problem-solving which is used to evaluate the difficulty of individual puzzles. In this paper, we elaborate further on the algorithmic hardness of Sudoku solving by studying the statistical mechanical properties of random ensembles of Sudoku instances. We show that Sudoku puzzles of varying sizes attain a distinctive peak in algorithmic hardness for a critical number of fixed cells, or \emph{clues}. This critical behavior is persistent throughout different solvers and reminiscent of the hardness of partial Latin square completion (PLS) \cite{gomes2002completing}, a closely related albeit easier and less constrained problem. We show that the hardness of random grids is consistent with that of real-world puzzles, showing that Sudoku publishers inadvertently design puzzles at the phase transition region.

Analogously to many hard combinatorial problems, the hardness peak of Sudoku solving is associated with a critical behavior in the constrainedness of random instances, which we uncover through linear relaxations and classification of backbone variables. In doing so, we provide the first description of a Sudoku \emph{freezing} transition, showing that the backbone fraction of random Sudoku grids undergoes a phase transition as the clue density is calibrated.

Sudoku is probably the most popular and widely-known $\mathcal{NP}$-Hard puzzle in present time. This puts it in a unique position amongst phase transition problems in which we can examine the statistical mechanical properties of elimination strategies employed by human solvers worldwide. We uncover a variety of critical phenomena in the applicability of Sudoku elimination techniques, which provide a method for calibrating the hardness of a puzzle in the design process. Building up on previous results about the Shannon entropy of complete Sudoku grids, we uncover yet another critical phenomenon in which the entropy of solutions to puzzles of varying clue density transitions between two phases at the point where the backbone size is at its largest.

\section{Problem Definition and Previous Work}

The Sudoku problem consists on completing a $n^2 \times n^2$ integer grid with numbers $1,2 \dots n^2$ such that all rows, columns and $n \times n$ blocks are filled without repetition. As a result, Sudoku can be seen as a set of $n^4$ variables subject to $3 n^2$ constraints of the \emph{alldifferent} type, whose study is of great relevance to the constraint programming community \cite{stergiou1999difference}. From the viewpoint of statistical mechanics, solving a Sudoku puzzle corresponds to finding the ground states of the following Hamiltonian \footnote{$\delta_{x,y} =
\left\{
    \begin{array}{ll}
        1  & \mbox{if } x = y \\
        0 & \mbox{if } x \neq y
    \end{array}
\right.$ is the Kronecker delta},

\begin{equation}
H(\sigma) = \sum_{i}\sum_{j \in \mathcal{N}(i)}{\delta_{\sigma_i,\sigma_j}}
\end{equation}

\begin{figure}[H]
\centering
\includegraphics[width=0.7\linewidth]{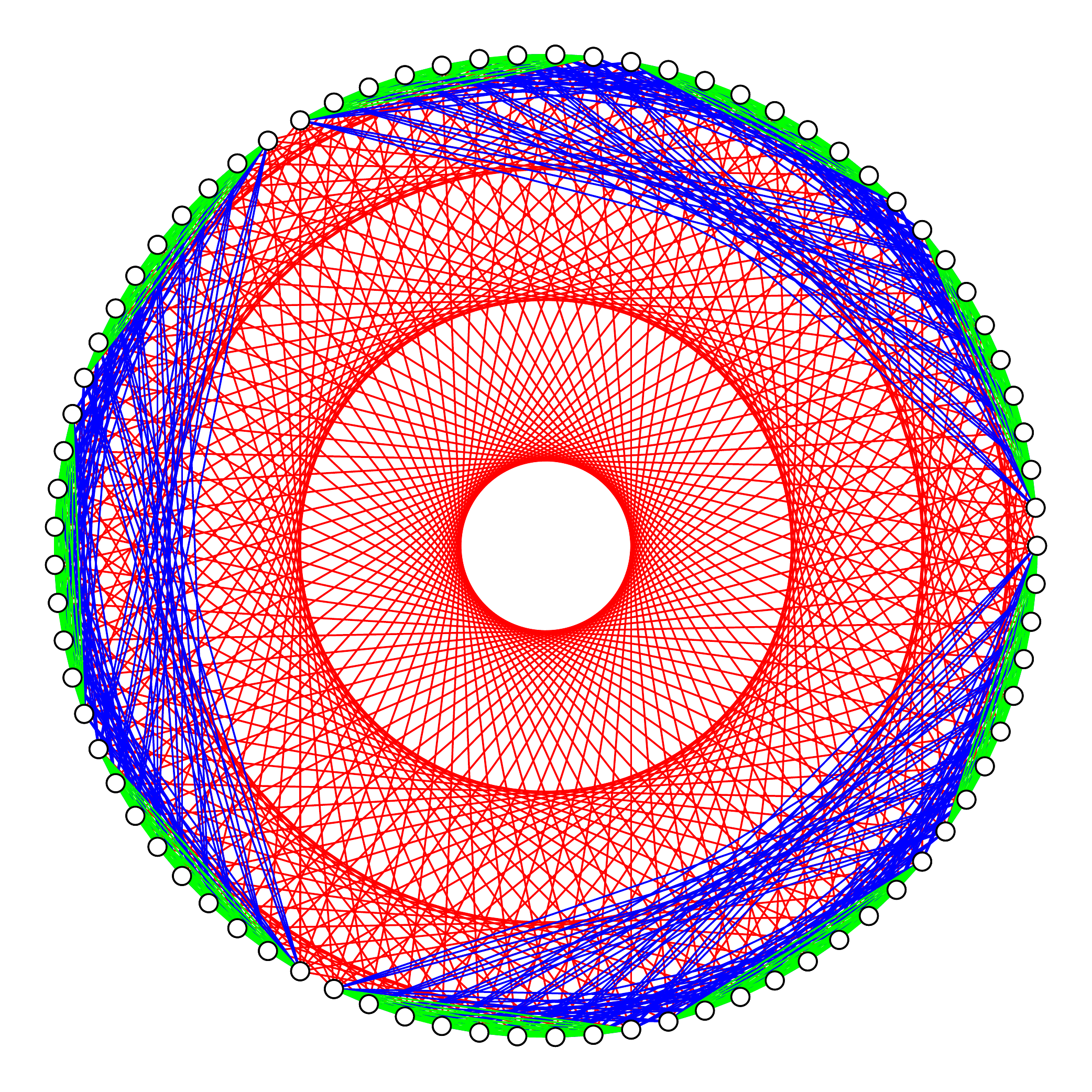}
\caption{Neigborhood graph for a $9 \times 9$ Sudoku. Every pair of neighbor cells is connected with an edge. Edges are painted according to the type of neighborhood relationship: red for \textbf{{\color{red!70!black}rows}}, green for \textbf{{\color{green!50!black}columns}} and blue for \textbf{{\color{blue}blocks}}.}
\label{fig:sudokuGraph}
\end{figure}

Where $i,j$ represent cells and $\mathcal{N}(i)$ denotes the neighborhood of a cell $i$ - that is, all cells which share a row, column or block with it. This is equivalent to a $n^2$-spin Potts hamiltonian with interations on a $3n^2 - 2n - 1$ neighbor periodic lattice (whose topology can be seen in Figure \ref{fig:sudokuGraph}) for a $9 \times 9$ Sudoku grid. The Sudoku Hamiltonian, albeit lacking a physical representation, provides valuable insights to the study of glassy systems, as \cite{williams2012paramagnetic} have shown.

\begin{figure}[H]
\centering
\includegraphics[width=0.7\linewidth]{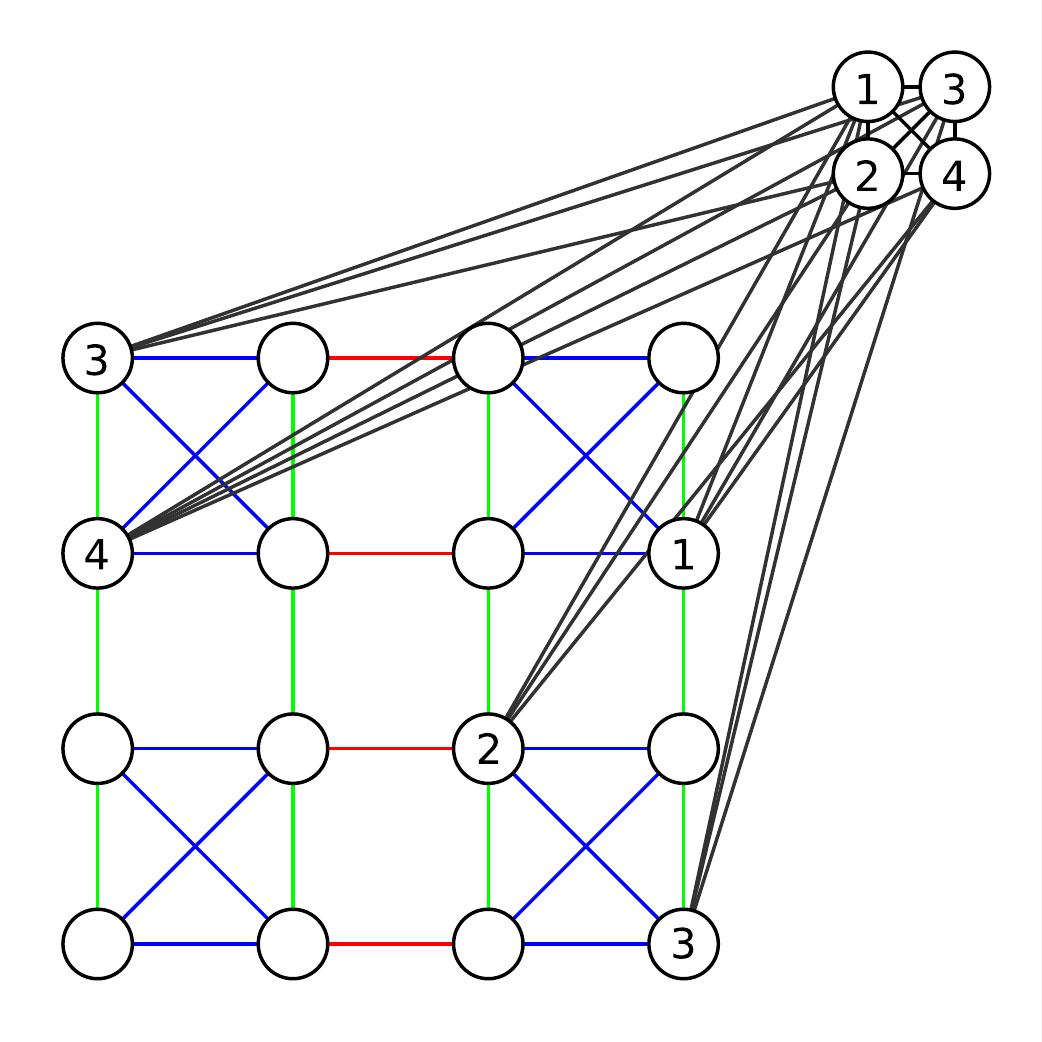}
\caption{$4 \times 4$ Sudoku puzzle recast as a graph coloring problem. Linking each clue cell to all vertices in the clique except that which corresponds to its value ensures a fixed coloring.}
\label{fig:sudokuGraphColoring}
\end{figure}

The topology $\mathcal{N}$ of cell neighborhoods can be used to translate Sudoku puzzles into instances of $k$-coloring, a widely relevant problem for which phase transitions are known to occur \cite{zdeborova2007phase,achlioptas2008algorithmic}. Recast in terms of graphs, each cell becomes a vertex, every two neighbor cells become connected by an edge and the Sudoku problem asks for a vertex coloring with at most $k=n^2$ colors such that no adjacent vertices have the same color, as Figure \ref{fig:sudokuGraphColoring} shows.

\section{Random instance generation}

Throughout this paper, we will examine critical phenomena on a random distribution of Sudoku grids which is produced by solving a blank grid with an exact solver. The solver's search parameters are randomized to produce different complete grids, and a randomly-chosen subset of cells are thereupon erased to produce a Sudoku instance with the desired clue density. The code used to produce these distributions is offered alongside with the code used to reproduce the experiments reported here in a Github repository (\url{https://github.com/marceloprates/Sudoku-Phase-Transitions}).

\section{Algorithmic Hardness of Sudoku Solving}

\begin{figure}[H]
\centering
\includegraphics[width=0.9\linewidth]{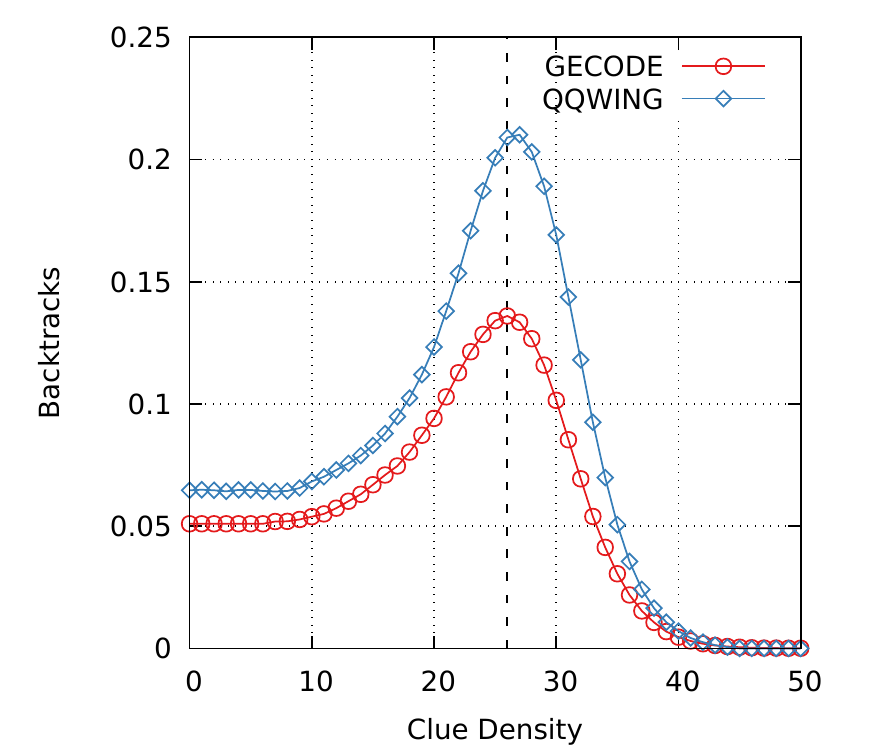}
\caption{The average number of backtracks required to solve ensembles of a hundred thousand random Sudoku puzzles attains a distinctive peak for a critical number of clues $\approx 26$ for both of the tested exact solvers. The dashed line marks the treshold of $26$ clues.}
\label{fig:gecode-qqwing-3}
\end{figure}

Hints of critical behavior in the performance of Sudoku solvers can be found in the literature, with \cite{cazenave2006search,lewis2007metaheuristics} reporting hindrances to their heuristic solutions at a specific number of pre-filled cells and \cite{williams2012paramagnetic} providing evidence of a frustrated, glassy state for puzzles with $\approx 27$ clues. These results are merely suggestive, as they are highly dependent on the employed heuristics and in the case of \cite{williams2012paramagnetic} talk more about the frustration of the Ising-like Sudoku lattice than about the performance of a real-world solver. If we want to uncover critical phenomena in the algorithmic hardness of Sudoku solving, a better approach is to solve ensembles of random puzzles of varying clue densities with an exact solver, whose performance can be measured deterministically by its backtrack count. We compare the average backtrack count of two different algorithmic approaches to Sudoku solving: a constraint programming solution defined over alldifferent constraints on the \emph{GECODE} CSP solver \cite{schulte2013view}, and \emph{QQwing} \cite{qqwing}, an open-source Sudoku solver built on top of problem-specific elimination strategies. Figure \ref{fig:gecode-qqwing-3} shows that for both solvers the algorithmic hardness (as measured by the average backtrack count) undergoes a peak at approximately the same location when solving ensembles of a hundred thousand random instances per clue density (GECODE peaks at $26$ and QQwing at $27$ clues). This result is consistent with the findings of \cite{cazenave2006search,lewis2007metaheuristics} and most interestingly with those of \cite{williams2012paramagnetic}, thus providing evidence that the statistical mechanics of glassy systems and the computational complexity of Sudoku are indeed linked.

\begin{figure}[H]
\centering
\includegraphics[width=0.9\linewidth]{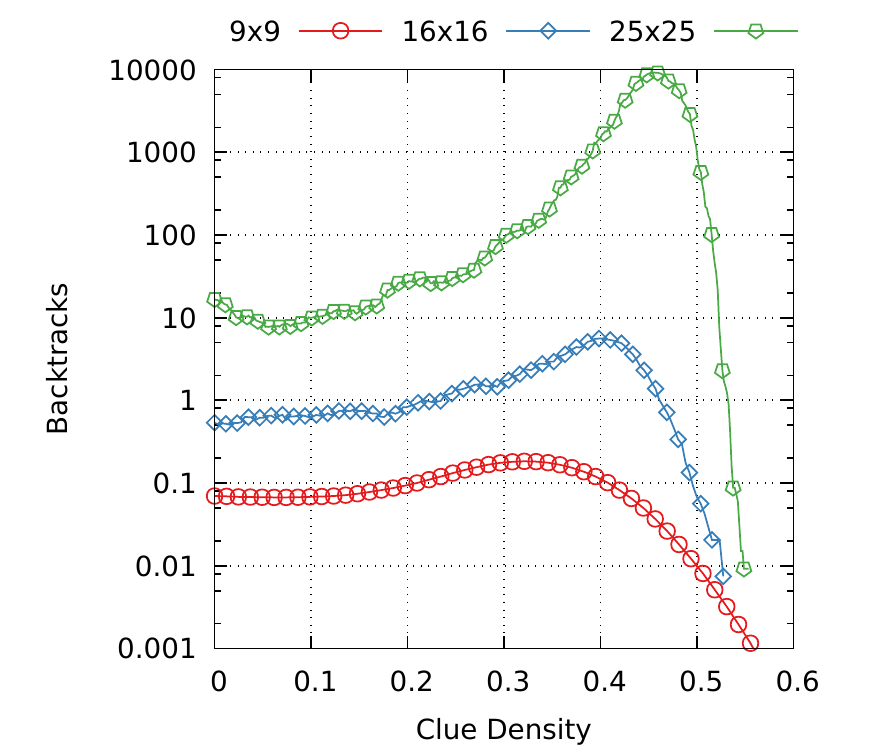}
\caption{As expected, the critical behavior extends to larger $16 \times 16$ and $25 \times 25$ puzzles. The theoretical exponential complexity associated with solving $n^2 \times n^2$ Sudokus is manifest in the peaks of the three hardness curves, plotted here on a logarithmic scale.}
\label{fig:qqwing-hardness-logscale}
\end{figure}

The critical behavior extends to larger puzzles, which yield higher and sharper peaks. We computed hardness curves averaged over ensembles of $100$ random instances per clue density for $9 \times 9$, $16 \times 16$ and $25 \times 25$ puzzles. They are shown in Figure \ref{fig:qqwing-hardness-logscale}, where the expected exponential complexity of Sudoku solving becomes manifest in the peaks' heights. The evolution of the peak locations also seems to suggest finite size scaling \cite{marino2016backtracking}, which is expected for phase transitions in combinatorial problems.

\begin{figure}[H]
\centering
\includegraphics[width=0.9\linewidth]{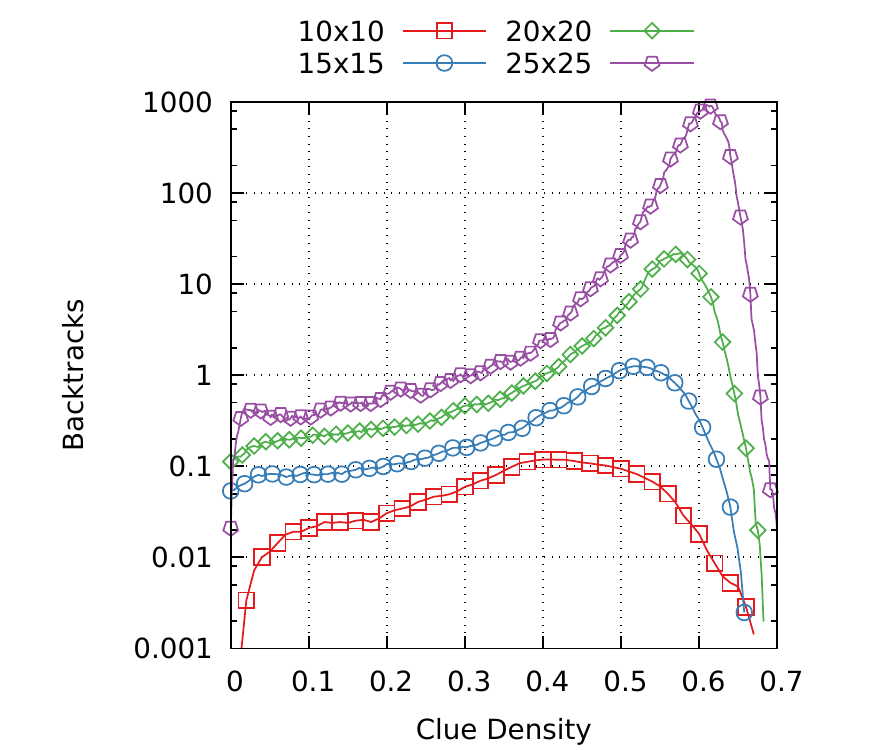}
\caption{The closely related problem of partial Latin square completion yields similar hardness curves, plotted here on a logarithmic scale for $n \times n$ grids of varying sizes.}
\label{fig:gecode-latinsquares}
\end{figure}

\begin{figure}[H]
\centering
\includegraphics[width=0.9\linewidth]{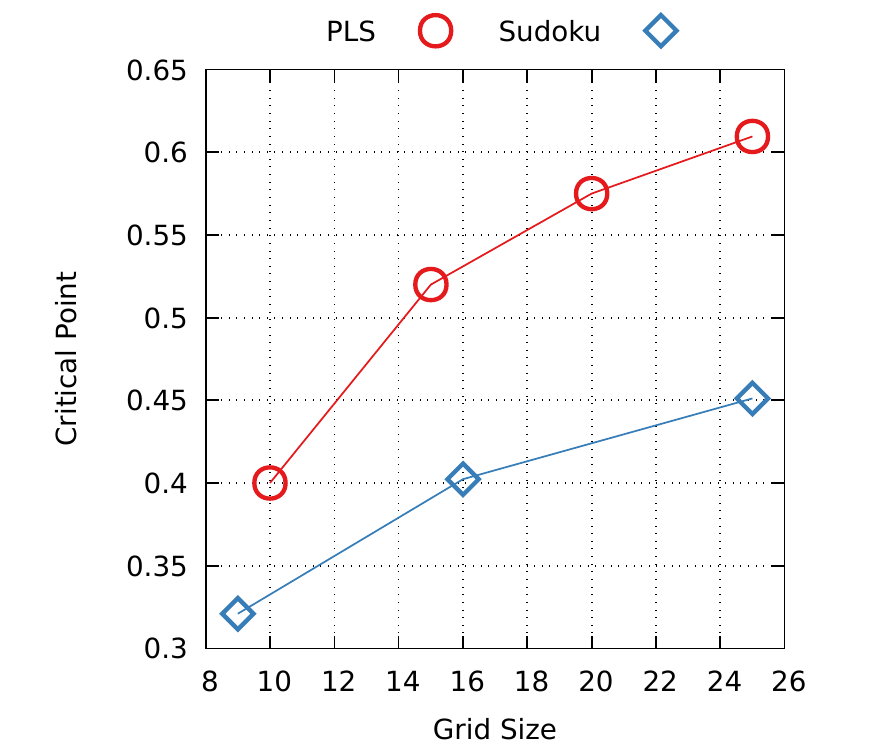}
\caption{The additional constrainedness of Sudoku rules is manifest in the location of its critical points, with random Sudoku puzzles becoming critically constrained before the corresponding partial Latin squares of the same size.}
\label{fig:criticalpoints-PLS-sudoku}
\end{figure}

It is insightful to analyze the hardness of Sudoku solving in comparison with that of partial Latin square completion (PLS), a simpler problem whose critical phenomena have received some attention in the scientific literature \cite{gomes2002completing}. We computed hardness curves averaged over ensembles of $100$ random instances for each clue density. Figure \ref{fig:gecode-latinsquares} plots the hardness curves for PLS on a logarithmic scale, in which one can identify critical phenomena similar to those observed for Sudoku solving. PLS instances are much easier, requiring at most one tenth of the backtracking compared to Sudoku puzzles of the corresponding size, as the data for $10 \times 10$, $15 \times 15$ and $25 \times 25$ grids shows. Additionally, because PLS instances are subject to a reduced number of alldifferent constraints we expect them to become overconstrained for larger clue densities than the corresponding Sudoku grids, as Figure \ref{fig:criticalpoints-PLS-sudoku} shows.

\begin{figure}[H]
\centering
\includegraphics[width=0.8\linewidth]{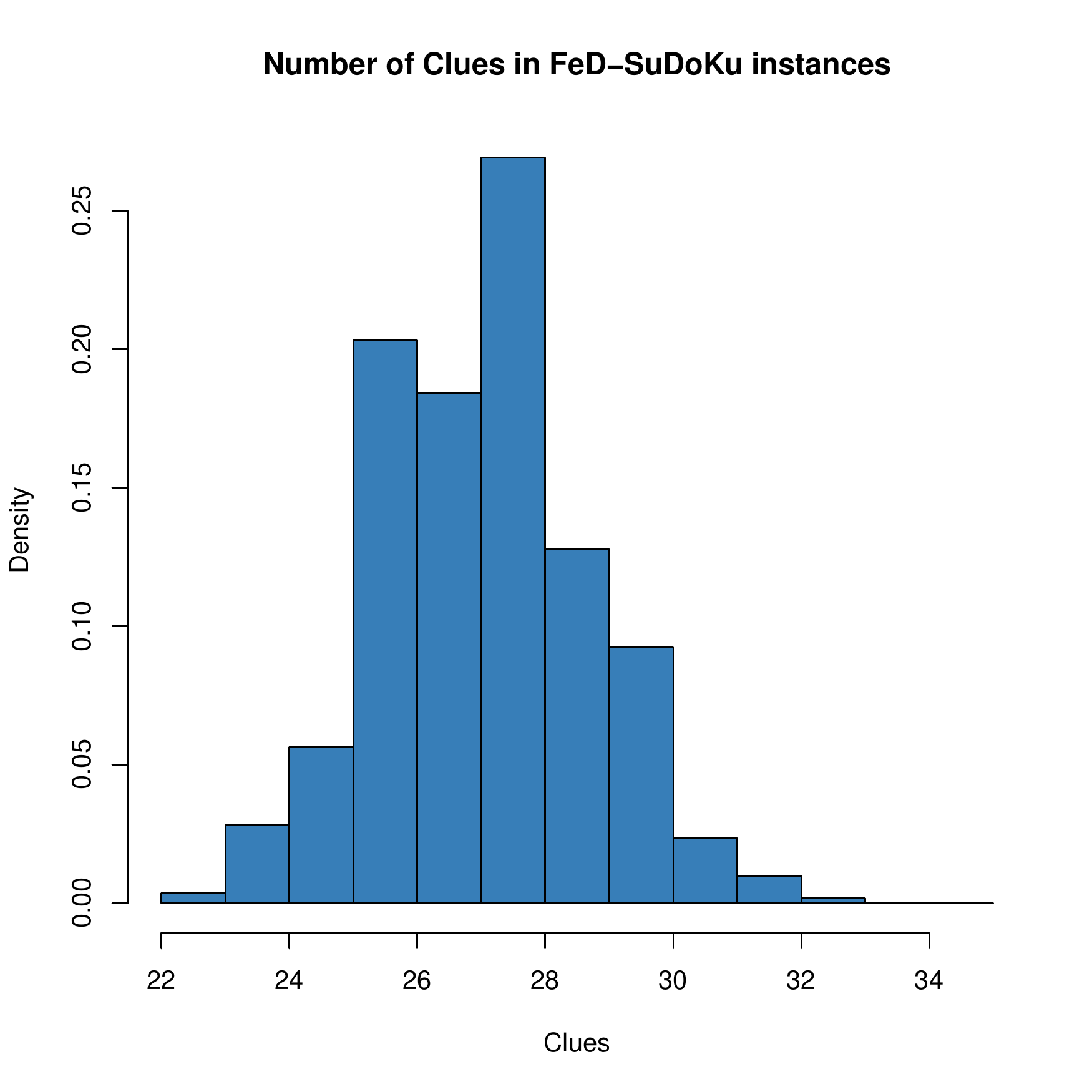}
\caption{Distribution of number of clues for puzzles in the fed-sudoku.eu website.}
\label{fig:clues-hist-fedsudoku}
\end{figure}

\begin{figure}[H]
\centering
\includegraphics[width=0.9\linewidth]{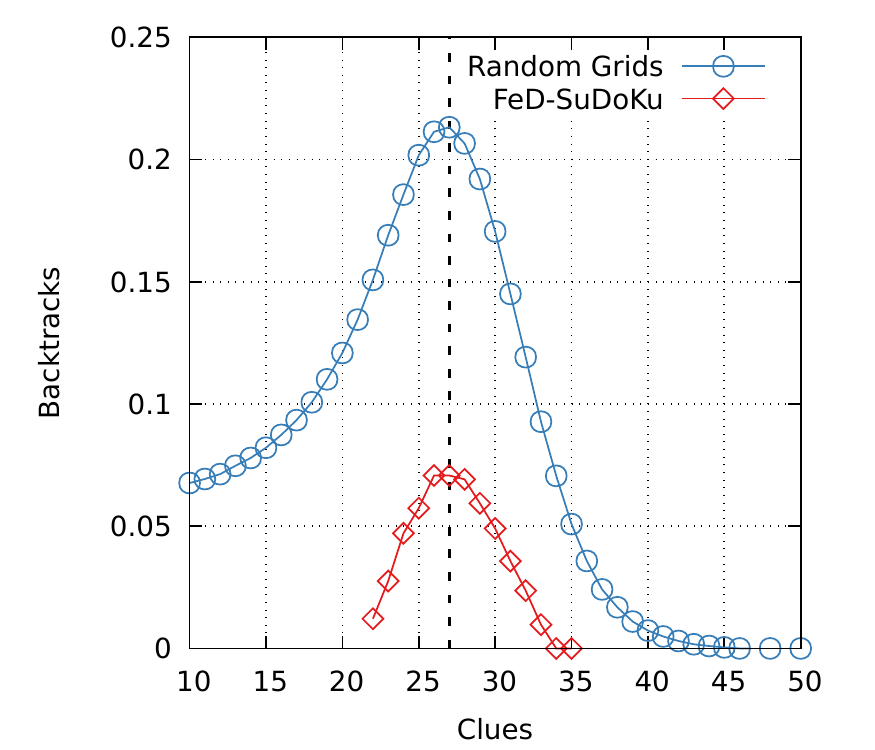}
\caption{Although not uniformly distributed among clue densities and much easier to solve, typical published puzzles capture the critical behavior of random Sudoku grids. The figure compares the average backtrack count of random Sudoku to that of puzzles in the fed-sudoku.eu website. The dashed line indicates the critical point of $27$ clues.}
\label{fig:fedsudoku-random-3}
\end{figure}

Published Sudoku puzzles are typically not equally distributed among different clue densities. Underconstrained instances with more than one solution as well as overconstrained instances leading to easy elimination techniques are typically filtered from such datasets, as Figure \ref{fig:clues-hist-fedsudoku} shows. Sudoku publishers, as a result, unintentionally design puzzles at the phase transition region, where the signature hardness of Sudoku solving can be harnessed. Figure \ref{fig:fedsudoku-random-3} compares the algorithmic hardness of random Sudoku to that of instances in the fed-sudoku.eu website, showing that both ensembles yield a hardness peak at the same critical point of $27$ clues.

\section{Linear Relaxations, Backbones and the Constrainedness of Random Sudoku Instances}

The connection between algorithmic hardness and constrainedness for hard computational problems has been known for at least twenty years \cite{cheeseman1991really,mertens2006easiest,dudek2017combining,bapst2016condensation}. When the problem space is bisected into underconstrained and overconstrained regions, it is easy to see that instances from both sets can be solved with relative ease. Underconstrained instances allow for a high density of solutions so that finding one is easy, while overconstrained instances give little trouble to backtracking algorithms as unsuccessful solution paths can be cut off early in the search process. Search methods face severe hindrances in the boundary between these two regions, which teems with instances with an exponential number of local optima separated by high energy barriers \cite{achlioptas2008algorithmic}. For a large number of $\mathcal{NP}$-Hard problems, it has been shown that this boundary is \emph{sharp}, meaning that statistical variables such as the satisfiability probability of a random k-SAT formula undergo abrupt transitions between the two phases as a specific set of control parameters is calibrated. Finite size instances deviate slightly from the general trend of a solution space with a discontinuous transition between under and overconstrainedness, in analogy to physical systems such a spin glass, for which the transition temperature between magnetic and non-magnetic behavior becomes discontinuous at the thermodynamic limit \cite{sherrington1975solvable}.

Random Sudoku grids lack a satisfiability transition because valid puzzles guarantee at least one solution, but their constrainedness can nevertheless be assessed. \cite{schawe2016phase} evaluate the constrainedness of the traveling salesperson problem by computing the probability that ensembles of instances yield integer solutions when solved with a linear relaxation of a integer programming model. The authors uncover a phase transition phenomenon in the probability $p(\sigma)$ that the LP relaxation is integer as a function of a disorder parameter $\sigma$ of Euclidean TSP instances. We can solve Sudoku puzzles with the ILP formulation proposed below and understand the hardness peaks of Figure \ref{fig:gecode-qqwing-3} in terms of the transition points of the LP relaxation curves shown in Figure \ref{fig:linrelax}. To produce these curves, we averaged the number of integral solutions over ensembles of $100$ random instances for each clue density.

\begin{figure}[h]
\begin{equation} \label{eq:sudokuILP}
\begin{array}{ll@{}ll}
\text{minimize}  & 0 &\\
\text{subject to}	& \displaystyle\sum\limits_{k=1}^{n^2}{x_{ijk}} = 1,~  &\substack{i=1 \dots n^2, \\ j=1 \dots n^2}\\
					& \displaystyle\sum\limits_{i=1}^{n^2}{x_{ijk}} = 1,~  &\substack{j=1 \dots n^2, \\ k=1 \dots n^2}\\
					& \displaystyle\sum\limits_{j=1}^{n^2}{x_{ijk}} = 1,~  &\substack{i=1 \dots n^2, \\ k=1 \dots n^2}\\
					& \displaystyle\sum\limits_{i=1}^{n}\sum\limits_{j=1}^{n}{x_{(a+i)(b+j)k}} = 1,~  &\substack{a=1 \dots n, \\ b=1 \dots n, \\ k=1 \dots n^2}\\
                 	& x_{ijk} \in \{0,1\},~ &\substack{i=1 \dots n^2, \\ j=1 \dots n^2, \\ k=1 \dots n^2}\\
\end{array}
\end{equation}
\end{figure}

\begin{figure}[h]
    \includegraphics[width=0.9\linewidth]{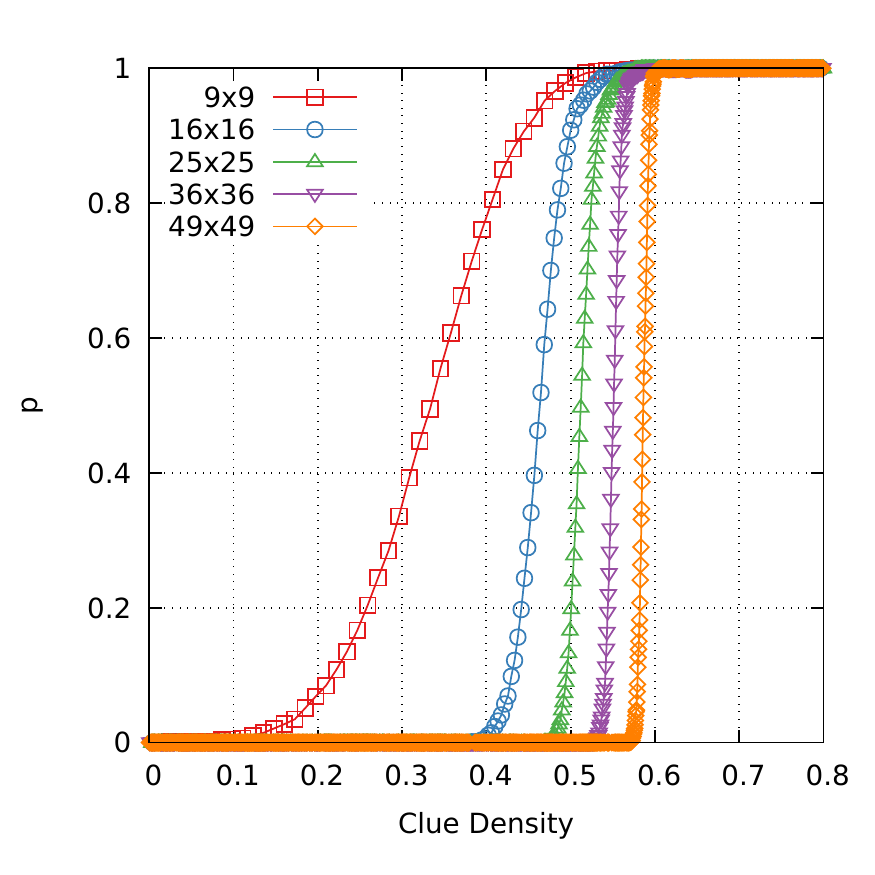}
    \caption{The probability of integral solutions in the linear relaxation of the integer linear problem of Equation \ref{eq:sudokuILP} undergoes interesting critical behavior as the clue density is calibrated, although the data does not suggest finite size scaling.}
    \label{fig:linrelax}
\end{figure}

\ifx
\begin{figure}[H]
    \includegraphics[width=0.9\linewidth]{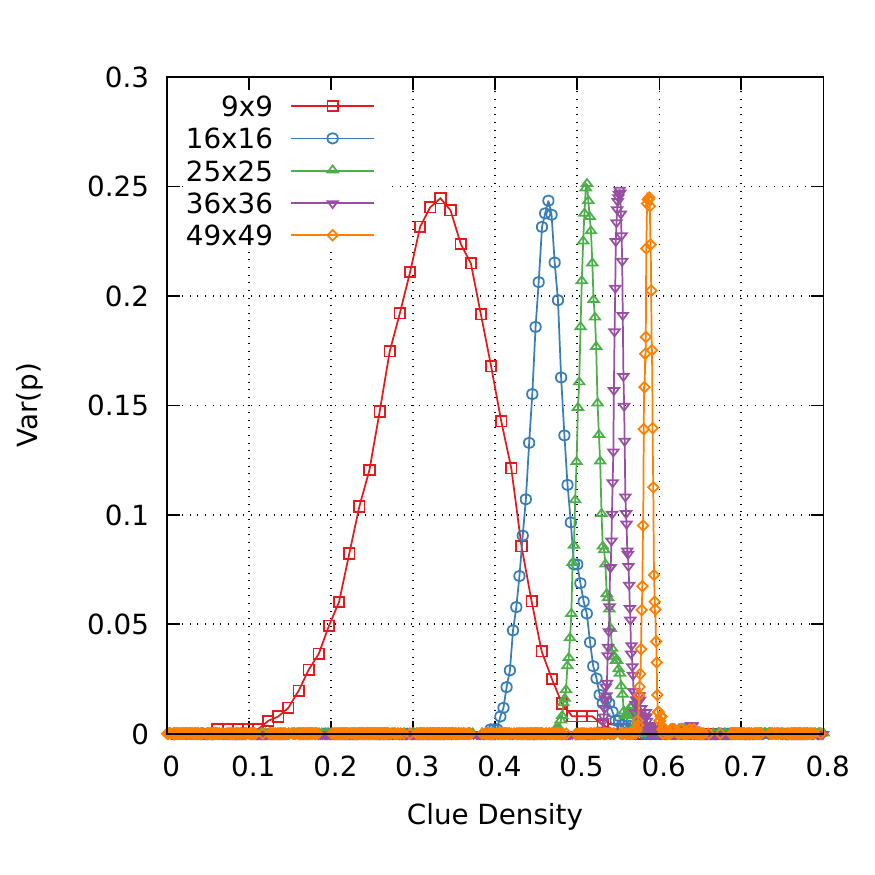}
    \caption{Variance of the probability that the LP relaxation is integer as a function of the clue density of ensembles of $100$ puzzles of varying sizes.}
    \label{fig:linrelax-var}
\end{figure}
\fi

One problem with the LP relaxation approach is its algorithm-dependency. If the ILP formulation were to change, the results in Figure \ref{fig:linrelax} could possibly change accordingly. This may help explain why the transition curves do not seem to suggest finite size scaling, as the LP approach possibly does not reflect the constrainedness of Sudoku instances to their full extent.

\begin{figure}[h]
\centering
\includegraphics[width=1\linewidth]{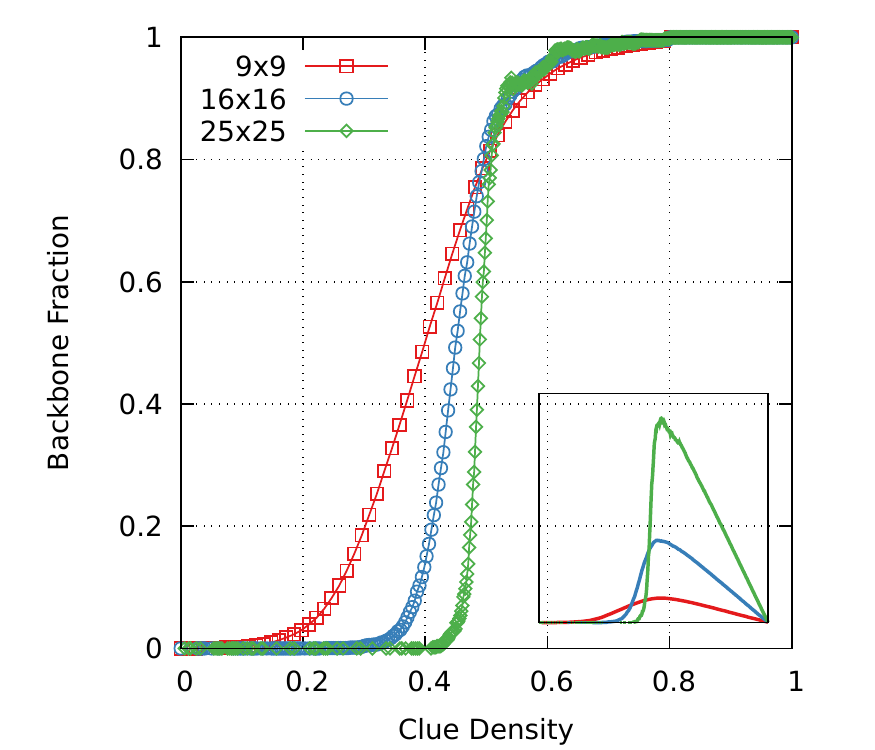}
\caption{Variation of the backbone fraction with the clue density of random ensembles of a hundred $9 \times 9$, $16 \times 16$ and $25 \times 25$ puzzles. The inset shows the respective backbone sizes.}
\label{fig:backbone-fractions}
\end{figure}

One way to evaluate the constrainedness of Sudoku puzzles whilst preserving algorithm independency is to look at its \emph{backbone}. The backbone of an instance is given by the set of its variables which are \emph{frozen}, meaning they assume the same value in all solutions. Large backbones impose hardship to search methods, and as a result the backbone size is closely linked to algorithmic hardness. Random k-SAT is known to undergo a \emph{freezing} phase transition just before the more famous satisfiability transition \cite{gogioso2014aspects}. \cite{pelanek2014difficulty} have provided evidence that published Sudoku puzzles become \emph{unfrozen} when a critical number of constraints are relaxed (i.e. the connectivity of the Sudoku graph in Figure \ref{fig:sudokuGraph} is decreased). Here we show that the backbone also undergoes critical behavior when the puzzles' clue density is calibrated. The inset of Figure \ref{fig:backbone-fractions} shows the variation in the backbone size for $9 \times 9$, $16 \times 16$ and $25 \times 25$ instances, which undergoes an exponential increase followed by a linear descent near the critical point.

Because the dimensionality of the Sudoku search space is given by the number of non-fixed cells, the true measure of its constrainedness is given by the \emph{backbone fraction}: the number of backbone variables per non-fixed cell. Figure \ref{fig:backbone-fractions} suggests that the backbone fraction undergoes a phase transition, which means instances become overconstrained after a critical number of clues is traversed. The increasing sharpness and horizontal shift of curves corresponding to larger puzzles also suggests scaling behavior. If finite size scaling applies, then the critical clue density $k_c(N)$ for finite size instances deviates from the critical clue density $k_c$ at the thermodynamic limit according to a power-law relationship given by $k_c(N) \sim k_c - \alpha \times N^{-1/\upsilon}$ for a positive constant $\alpha$ and a critical exponent $\upsilon$, although we were unable to obtain enough data to compute these parameters with precision.

\ifx
\begin{figure}[h]
\centering
\includegraphics[width=0.8\linewidth]{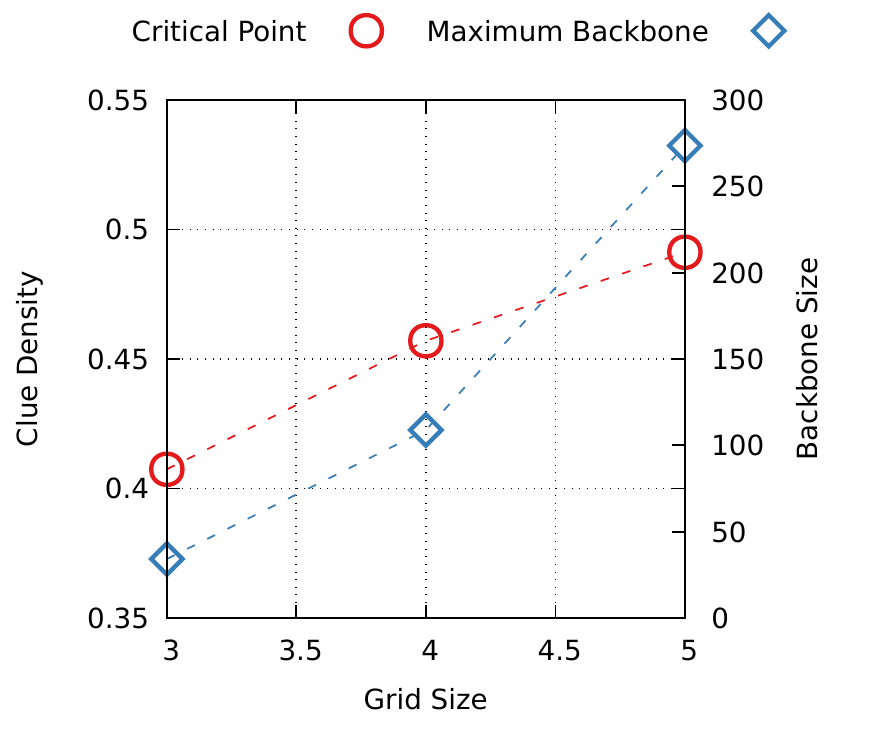}
\caption{The evolution of the critical clue density exhibits signs of scaling behavior. For large $N$, the critical point $k_c(N)$ is expected to converge to $k_c$, the critical clue density at the thermodynamic limit. Also for large $N$, the maximum backbone size (plotted on the secondary axis) is expected to converge to $k_c(N) N^4$.}
\label{fig:transition-points}
\end{figure}
\fi

\section{Criticality of Sudoku Solving Strategies}

\begin{figure}[h]
\centering
\includegraphics[width=1\linewidth]{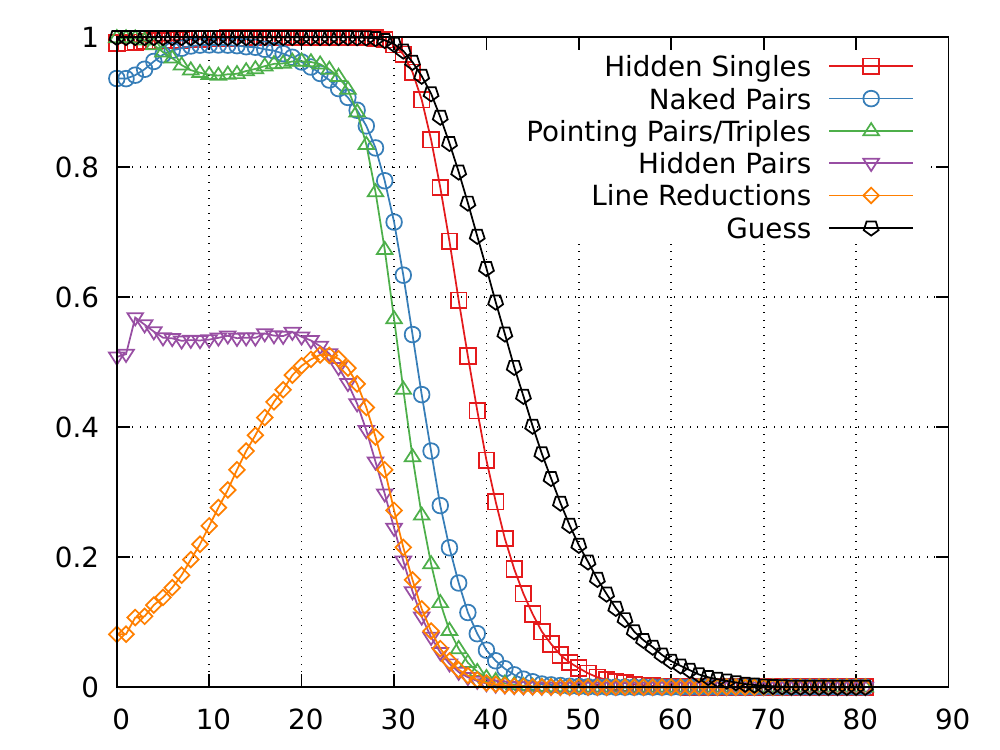}
\caption{Variation in the application frequency of different Sudoku elimination strategies as a function of the number of clue cells for random ensembles of a hundred thousand $9 \times 9$ puzzles.}
\label{fig:strategies-prob-3}
\end{figure}

\begin{figure}[h]
\centering
\includegraphics[width=1\linewidth]{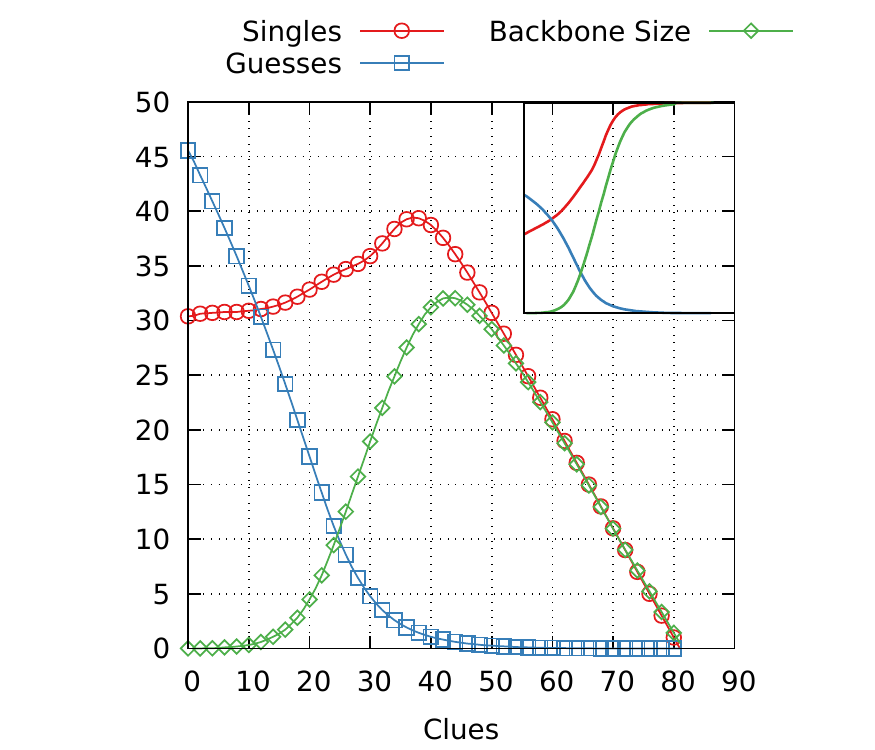}
\caption{Variation in the frequency of singles, guesses and backbone size as a function of the number of clue cells for random ensembles of a hundred thousand $9 \times 9$ puzzles. The inset shows the fraction of singles, guesses and backbone variables per non-fixed site.}
\label{fig:singles-guesses-backbone-3}
\end{figure}

Phase transitions and statistical mechanics offer valuable insights into the average algorithmic difficulty of $\mathcal{NP}$-Hard problems, but in our case we can do even more. Sudoku's unique position as a universally recognized $\mathcal{NP}$-Hard problem enables us to study the statistical mechanics of elimination strategies employed by human solvers worldwide, bridging the worlds of average algorithmic hardness and human problem-solving. Figure \ref{fig:strategies-prob-3} shows how the application frequency of different elimination strategies varies with the puzzles' clue density, and in doing so offers a glimpse at many critical phenomena. Hidden singles, naked pairs and pointing pairs/triples are applied with close to $1$ probability for underconstrained, low clue density puzzles, but undergo a sharp transition as the clue density is increased and the simplest strategy - singles - increasingly suffices to solve most puzzles (see Figure \ref{fig:singles-guesses-backbone-3}). The frequency of hidden pairs, which starts at $\approx 0.51$, follows the same critical behavior. Interestingly, the frequency of block-line reductions peaks at $23$ clues. The guess probability also seems to undergo a critical transition from $1$ to $0$ as the increasing constrainedness of high clue density puzzles enables one to solve them by logic alone.

\begin{figure}[h]
\centering
\includegraphics[width=1\linewidth]{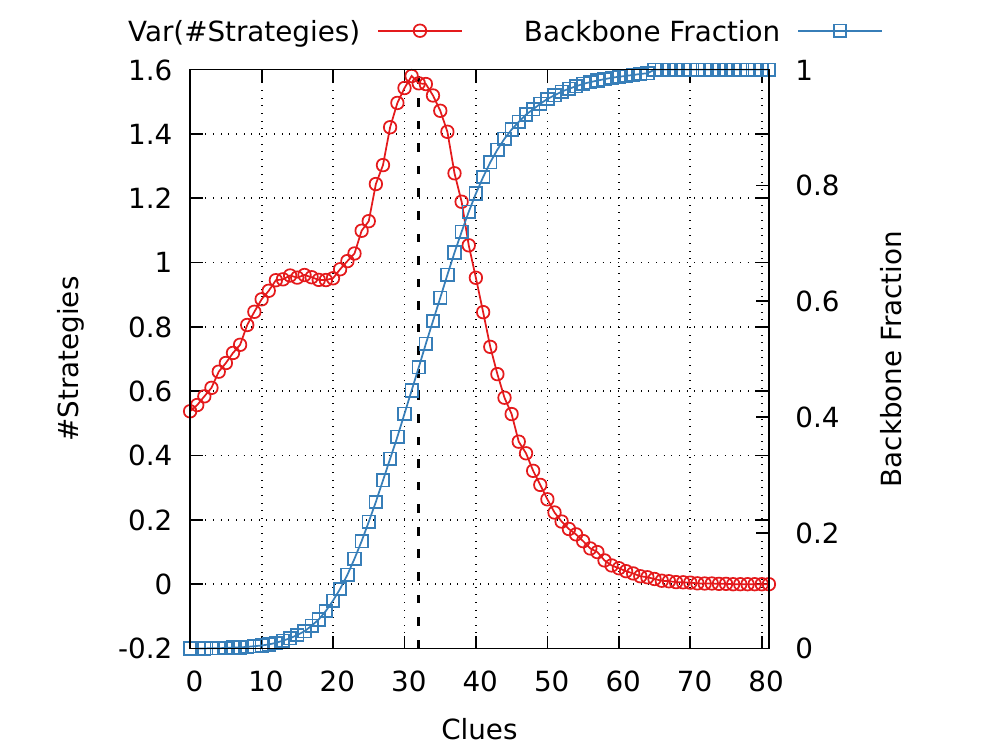}
\caption{Variance of the number of required strategies vs. backbone fraction for random ensembles of a hundred thousand $9 \times 9$ puzzles. The dashed line marks the transition point of the backbone fraction transition.}
\label{fig:strategies-uncertainty-backbone-3}
\end{figure}

Figure \ref{fig:singles-guesses-backbone-3} shows how the frequency of singles (non-fixed cells which admit a single possibility) relates to the backbone size. As the fraction of backbone variables approaches $1$, frozen variables become manifest in this elimination strategy, which enables one to solve puzzles just by repeatedly applying it. We may assume that the challenge of solving a Sudoku grid is linked to the variety of elimination strategies required to do so, which suggests that the ubiquity of singles is symptomatic of boring puzzles. In fact, Figure \ref{fig:strategies-uncertainty-backbone-3} shows that the uncertainty about the set of strategies required to solve a puzzle is at its maximum at the transition point of the backbone fraction transition, which adds credibility to this hypothesis.

\section{Shannon Entropy of Random Sudoku Grids}

Random Sudoku grids - that is, randomly selected $n^2 \times n^2$ matrices obeying the Sudoku alldifferent constraints - are complicated combinatorial objects, but their informational entropy \cite{6773024} offers valuable insights about the structure of the distribution. \cite{Newton2010} evaluate the change in the amount of disorder when a transition is made from random $n^2 \times n^2$ integer matrices to random Sudoku grids. In particular they show how the additional constraints imposed by Latin squares and Sudoku grids compared to integer matrices sparkle an increase in their informational entropy, which can be computed through the singular value decomposition of the component-wise average of ensembles of Sudoku grids.

For an ensemble $X = \{X_1, X_2, \dots X_N\}$ of $n^2 \times n^2$ Sudoku matrices, we can compute the component-wise average $A$ and the singular values $\sigma_i$ are defined as $\sigma_i = \sqrt{\lambda_i}$, where $\lambda_i$ are the eigenvalues of the covariance matrix $A^{T} A = \lambda_i \vec{v_i}$. We also introduce the normalized singular values $\hat{\sigma_i} = \sigma_i / \sum{\sigma}$. The Shannon entropy of an ensemble of $n^2 \times n^2$ Sudoku grids is then given by $H = - \sum_{i=1}^{n^2}{\hat{\sigma_i} \ln{\hat{\sigma_i}}}$.

\begin{figure}[H]
\centering
\includegraphics[width=1\linewidth]{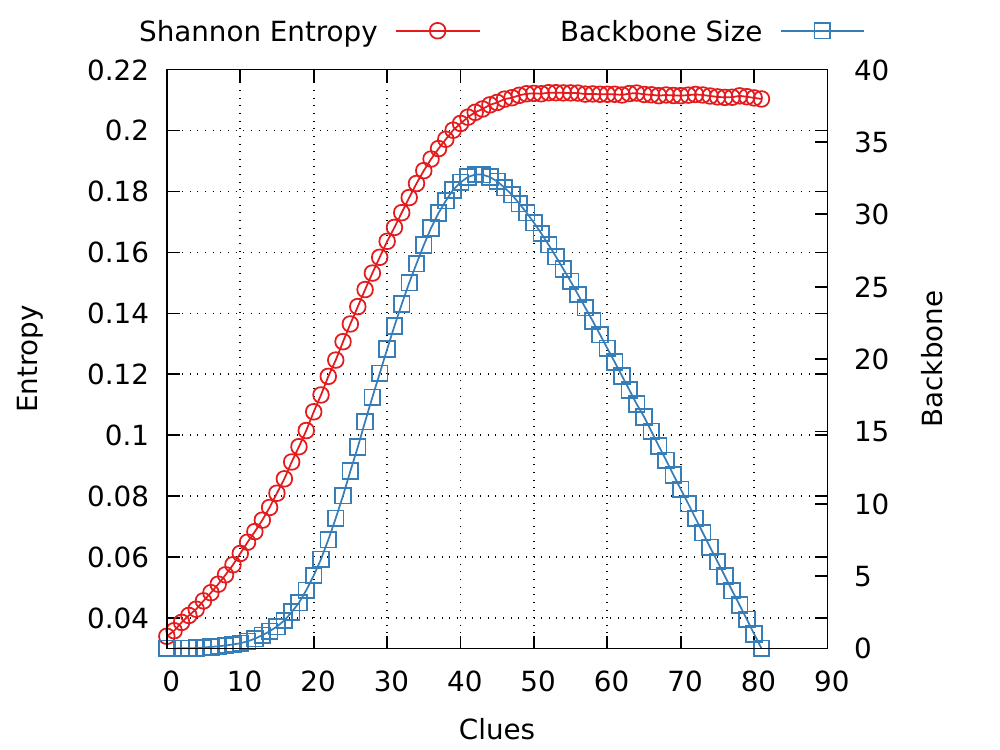}
\caption{Variation in the Shannon entropy of Sudoku solutions vs. backbone size as a function of the number of clue cells for random ensembles of a hundred thousand $9 \times 9$ puzzles.}
\label{fig:entropy-backbone-3}
\end{figure}

Random complete Sudoku grids are more constrained than the corresponding integer matrices of the same size, but when incomplete grids are considered, the puzzles' clue density introduces yet another spectrum of constrainedness. We can therefore expect the Shannon entropy of solutions to higher clue density puzzles to be larger than that of sparse grids. Figure \ref{fig:entropy-backbone-3} shows that this is indeed the case, and additionally{} that the Shannon entropy of Sudoku solutions appears to undergo a critical behavior, attaining a plateau close to the location where the backbone size is the largest. The interpretation of this result is that solutions to lower clue density puzzles are less \emph{biased}, in the sense that their average converges much faster to the unbiased average matrix $J_{n^2} \left( \frac{1}{n^2}
\sum_{i=1}^{n^2}{i}{} \right)$ \footnote{$J_{n}$ stands for the matrix of ones, the $n \times n$ square matrix where all elements are $1$.} as more elements are added.

The Shannon entropy interestingly plateaus at an intermediary point in the clue density spectrum, suggesting that the solution bias undergoes no significative changes after a critical point is traversed. This result is consistent with other critical phenomena presented in this paper, and particularly with the freezing transition shown in Figure \ref{fig:backbone-fractions}.

\ifx
\section{Nearest Neighbor Solutions, Random Walks and the Sudoku Clustering Transition}

The hardship faced by search algorithms in combinatorial problems has much to do with the statistical properties of the solution space structure, and in particular with its connectivity. It has been shown both theoretically and through simulations that random k-SAT formulas undergo a clustering phase transition as the clauses-to-variables ratio $M/N$ is increased. Initially, the set of solutions is concentrated on a single giant cluster of satisfying boolean assignments connected by low energy movements, such as a single variable flip. As we allow $M/N$ to increase and random formulas become more constrained, this single set shatters into exponentially many clusters separated from one another by high energy barriers, before finally condensating into a sub-exponential number of relevant ones. This clustering explosion is responsible for the frustration of search methods, which, unable to traverse the high energy barriers between clusters, become trapped in local optima.

It is easy to define a distance metric for k-SAT solutions. Because they are essentially binary strings, the Hamming distance

\begin{equation}
d(\sigma^1,\sigma^2) = \displaystyle\sum_{i=1}^{N}{\delta_{\sigma_i^1,\sigma_i^2}}
\end{equation}

is a natural candidate, and one can then analyze the connected components of solutions separated by nearest neighbor movements of Hamming distance $d(\sigma^1,\sigma^2) \leq 1$. This neighborhood function clearly does not work in the case of Sudoku, where single variable flips always translate a satisfying assignment into a non-satisfying one. What we can do instead is consider the neighborhood of solutions separated by special kinds of two-variable exchanges in the same unit (row, column or $n \times n$ block).
\fi

\section{Concluding Remarks and Future Works}

Sudoku is a widely popular $\mathcal{NP}$-Complete combinatorial puzzle whose prospects for studies in human reasoning/computation has been recently demonstrated, but the hardness of Sudoku solving has surprisingly received little attention. Hoping to shed light on this matter, in this paper we explore the statistical mechanics of random Sudoku instances. We show that in the likes of many difficult combinatorial problems, the hardness of random Sudoku is linked to critical behavior in the constrainedness of ensemble instances, which undergoes phase transition phenomena as the clue density is calibrated. In doing so, our paper provides the first description of a Sudoku freezing transition, in which the fraction of backbone variables abruptly shifts from zero to one when a threshold of pre-filled cells is traversed. These findings agree with the hardness of real-world instances, showing that Sudoku publishers inadvertently design puzzles at the phase transition region. Our results also extend the behavior of random partial Latin square completion when additional block constraints are added, and demonstrate how the added constrainedness renders instances harder. We demonstrate that the criticality of random Sudoku can also be seen in the disorder of ensembles of solved puzzles, whose Shannon entropy stops increasing after instances become critically constrained. Our results also uncover a variety of critical phenomena in the applicability of Sudoku elimination strategies employed by human solvers worldwide. Because Sudoku publishers typically rate puzzles according to which kind of elimination rules they require to solve, these results provide a method for calibrating the difficulty of a puzzle in the design process.

The findings presented in this paper shed light on the nature of the $k$-coloring transition for fixed graph topologies, and are an invitation to the study of phase transition phenomena in problems defined over \emph{alldifferent} constraints. They also suggest that there are advantages to studying the statistical mechanics of popular puzzles, as it enables one to further bridge computational complexity and human problem solving. For future works, we are interested on how the critical phenomena presented in this paper possibly relate to the clustering of solutions. A clustering transition, if found, would help explain the difficulty of random Sudoku instances by describing the high energy barriers which separate "islands" of solutions connected by small movements, such as the relabeling of two digits on a Sudoku board.

\section{Acknowledgments}

This study was financed in part by the Coordenação de Aperfeiçoamento de Pessoal de Nível Superior - Brasil (CAPES) - Finance Code 001 and the Conselho Nacional de Desenvolvimento Científico e Tecnológico (CNPq).

\bibliographystyle{plain}
\bibliography{main}

\end{document}